\documentclass[]{IEEEtran}
\usepackage{graphicx}
\usepackage{amsmath}
\usepackage{cite}
\usepackage{url}

\allowdisplaybreaks[4]

\begin{document}

\title{Exact Solution of the Full RMSA Problem in Elastic Optical Networks}

\author{Fabio David,  José F. de Rezende, Valmir C. Barbosa
\thanks{We thank Luidi Simonetti for many enlightening conversations on integer
programming and Gurobi usage. This work was supported in part by Conselho
Nacional de Desenvolvimento Científico e Tecnológico (CNPq), Coordenação de
Aperfeiçoamento de Pessoal de Nível Superior (CAPES), and grants from
Fundação Carlos Chagas Filho de Amparo à Pesquisa do Estado do Rio de Janeiro
(FAPERJ). This work was also supported by MCTIC/CGI.br/São Paulo Research
Foundation (FAPESP) through projects Slicing Future Internet Infrastructures
(SFI2) – grant number 2018/23097-3, Smart 5G Core And MUltiRAn Integration
(SAMURAI) – grant number 2020/05127-2 and Programmable Future Internet for
Secure Software Architectures (PROFISSA) – grant number 2021/08211-7.
\textit{(Corresponding author: Fabio David.)}}
\thanks{FD is with the Federal University of Rio de Janeiro, Informatics and
Computer Science Department (NCE). JFR and VCB are with the Federal University
of Rio de Janeiro, Systems Engineering and Computer Science Program, Centro de
Tecnologia, Sala H-319, 21941-914 Rio de Janeiro - RJ, Brazil (e-mail:
fabio@land.ufrj.br).}}

\maketitle

\begin{abstract}
Exact solutions of the Routing, Modulation, and Spectrum Allocation (RMSA)
problem in Elastic Optical Networks (EONs), so that the number of admitted
demands is maximized while those of regenerators and frequency slots used are
minimized, require a complex ILP formulation taking into account frequency-slot
continuity and contiguity. We introduce the first such formulation, ending a
hiatus of some years since the last ILP formulation for a much simpler RMSA
variation was introduced. By exploiting a number of problem and solver
specificities, we use the NSFNET topology to illustrate the practicality and
importance of obtaining exact solutions.
\end{abstract}


\section{Introduction}
\label{sec:intro}

Even though data transmission through optical fibers has been a reality for a
few decades, the constantly growing demand for higher capacity and greater
maximum reach has led research and development in the field to be always
evolving. One of the most promising technologies that emerged in the last decade
or so and is now particularly well poised to play an important role in the
coming years is based on the concept of an Elastic Optical Network (EON)
\cite{Jinno-6146481}. Like networks based on the more mature and still widely
used Dense Wavelength Division Multiplexing (DWDM) technology \cite{DWDM}, EONs
provide for the sharing of each link's spectrum between end-to-end demands, and
also for the use of so-called regenerators at certain nodes so that a signal
nearing its modulation's maximum reach can be regenerated and continue on its
designated end-to-end route. In both respects, however, EONs improve on DWDM
networks substantially. First, spectrum sharing in EONs is based on relatively
narrow, fixed-width frequency slots (FSs) that can be concatenated to provide
the demand with higher, reduced-waste capacity. Second, by employing modern
optical transceivers, the same link can carry demands using different
modulations (hence with different numbers of FSs if the demands have the same
bandwidth). This means that an EON regenerator can change the modulation used to
serve a demand and along with it the number of FSs used.

Regenerators are expensive and must therefore be used sparingly. The problem of
selecting the nodes at which to install them in a DWDM network is already a
difficult problem in its own right \cite{RLPgeneralized}. In EONs, however, this
gets considerably more complicated by the need to select not only which
modulation to use for each demand going through each regenerator but also which
FSs to allocate to it. In this case, an important concept is that of a segment,
which is a path between two nodes without a regenerator at any intermediate
node. Therefore, a path containing $R$ regenerators at intermediate nodes
comprises $R+1$ segments. While traversing any given segment, a demand uses the
same modulation and FSs on all the segment's links. Given the network topology
and a set of available modulations, each characterized by a bandwidth and a
maximum reach, one can readily enumerate all possible segments by considering
all paths on the graph. Given a set of demands, deciding how to route them using
these segments, and consequently how many regenerators to deploy and where, is
the NP-hard problem we deal with in this letter, known as the Routing,
Modulation, and Spectrum Allocation (RMSA) problem
\cite{Chatterjee-7105364}.

We continue by describing the relevant state of the art and our contribution in
Section~\ref{sec:contrib}, then in Sections~\ref{sec:form1} and~\ref{sec:form2}
introduce two ILP formulations for RMSA that for the first time consider some
of the most relevant objectives in EONs while abiding by every constraint they
impose. We present computational results and conclude in
Section~\ref{sec:results}.

\section{State of the Art and Contribution}
\label{sec:contrib}

In EONs, FS concatenation inside a segment requires that the FSs be perfectly
aligned between successive links and, on each link, that they be contiguous.
These constitute a continuity and contiguity (CC) criterion that must translate
into constraints in any RMSA formulation. With these and other constraints in
place, ideally multiple objectives should be pursued, including admitting as
many demands as possible while globally using as few regenerators and FSs as
possible. To the best of our knowledge, the single previous attempt to provide
the RMSA problem with a formulation for exact solution is the one in
\cite{turco}, which is an ILP formulation without CC constraints that targets
essentially the minimization of the total number of regenerators to be deployed.

The formulation we introduce, called RMSA-BP to emphasize the user-centric goal
of minimizing the chance that a demand is blocked, targets all three objectives
and includes CC constraints. We use the NSFNET topology to demonstrate the
practical feasibility of obtaining exact solutions, as well as to demonstrate
potential benefits in analyzing critical network properties.

\section{RMSA-BP Formulation}
\label{sec:form1}

We represent the network by an undirected graph $G$ of vertex set $N$ and edge
set $E$, where $N$ is the set of network nodes and $E$ is the set of network
links. Every link has the same link capacity (LC), given by how many frequency
slots (FSs) it has. We use $P$ to denote the set of segments, with segment
$p\in P$ beginning at node $s_p$ and ending at node $t_p$, and $I_p$ to denote
the first link on segment $p$. The set of demands is denoted by $D$, with demand
$d\in D$ being characterized by its node of origin $S_d$, its node of
destination $T_d$, and a bandwidth $B_d$. The number of FSs corresponding to
demand $d$ on segment $p$ is denoted by $F_p^d$ and given by the ratio of $B_d$
to the bandwidth provided by each FS for the modulation used in $p$. We use
$R_\mathrm{max}$ to denote the maximum number of regenerators that can be used
per demand. Notably, even though links in $E$ are undirected, each segment in
$P$ is inherently directed. Thus, a link can be traversed by a segment in either
of the link's two directions.

RMSA-BP is stated in terms of one set of constants, three sets of variables, and
also some shorthands that allow for a cleaner exposition. We give the full
formulation next, but divide the subsequent discussion into two separate parts,
depending on which variables, shorthands, and constraints relate to the
problem's CC requirement. We use weights $w_1\gg w_2\gg w_3$ in the objective
function only to symbolize that the problem's three goals are to be prioritized
in the following order: 1) Maximize the number of admitted demands; 2) Minimize
the total number of regenerators used; 3) Minimize the total number of FSs used.
The specific technique we use to enforce this prioritization is discussed in
Section~\ref{sec:results}.
\begin{align}
\text{max }
& \sum_{d\in D}\left(w_1a_d-w_2R_d-w_3F_d\right)
\nonumber\\
\text{s.t. }
\nonumber\\
& x_p^d\le a_d
& \forall d\in D, p\in P \tag{C1}\label{C1} \\
& C_n^d=\begin{cases}
a_d &\text{if }n=S_d \\
-a_d &\text{if }n=T_d \\
0 &\text{otherwise}
\end{cases}
& \forall d\in D, n\in N \tag{C2}\label{C2} \\
& R_d\le R_\mathrm{max}
& \forall d\in D \tag{C3}\label{C3} \\
& z_d^e\le X_d^e\;\text{LC}
& \forall d\in D, e\in E \tag{C4}\label{C4} \\
& z_d^e+F_d^e-X_d^e\le\text{LC}
& \forall d\in D, e\in E \tag{C5}\label{C5} \\
& X_d^eY_p^ez_d^e-X_d^{I_p}Y_p^{I_p}z_d^{I_p}=0
& \forall d\in D, p\in P, \tag{C6}\label{C6} \\
&& e\in p, e\neq I_p
\nonumber\\
& \sum_{\genfrac{}{}{0pt}{}{d,d'\in D}{d\neq d'}}O_{d,d'}^e=0
& \forall e\in E \tag{C7}\label{C7}
\end{align}

For each $p\in P$ and each $e\in E$, we use $Y_p^e$ to indicate whether link $e$
is part of segment $p$. In the affirmative case, the direction of traversal of
$e$ by $p$ is given implicitly by nodes $s_p$ and $t_p$. The $Y_p^e$'s are
binary constants whose values (from $\{0,1\}$) are assigned along with the
determination of the network's set of segments $P$ (see
Section~\ref{sec:intro}). The essential variables for use when the CC
requirement is disregarded are all binary as well and are grouped into two
sets: $a_d$, for each $d\in D$, indicating whether demand $d$ is admitted; and
$x_p^d$, for each $d\in D$ and each $p\in P$, indicating whether demand $d$ uses
segment $p$. These variables are sometimes used directly in the above
formulation, and also sometimes indirectly through the following shorthands:
\begin{align}
R_d &= \sum_{p\in P}x_p^d-a_d,
& F_d^e &= \sum_{p\in P}F_p^dx_p^dY_p^e,
\nonumber\\
F_d &= \sum_{e\in E}F_d^e,
& C_n^d &=
\sum_{\genfrac{}{}{0pt}{}{p\in P}{s_p=n}}x_p^d-
\sum_{\genfrac{}{}{0pt}{}{p\in P}{t_p=n}}x_p^d.
\nonumber
\end{align}
In these equations, $R_d$ is the number of regenerators used by demand $d$,
$F_d^e$ is the number of FSs used by demand $d$ on link $e$, $F_d$ is the total
number of FSs used by demand $d$, and $C_n^d$ is the flow deficit at node $n$
for demand $d$ (that is, the number of segments used by $d$ that are outgoing
from $n$ in excess of those that are incoming to $n$). If $x_p^d=1$ for at most
one segment $p$, then $C_n^d\in\{-1,0,1\}$.

Only Constraints~(\ref{C1})--(\ref{C3}), along with
$\sum_{d\in D}F_d^e\le\text{LC}$ for each $e\in E$, henceforth called
Constraint~(Cx) for the sake of the argument, are needed if CC need not hold. Of
these, Constraints~(\ref{C1}) and~(\ref{C2}) take care of how the $a_d$'s and
the $x_p^d$'s relate to one another, as follows. A blocked demand $d$ ($a_d=0$)
uses no segments ($x_p^d=0$ for every $p\in P$). An admitted demand $d$
($a_d=1$), on the other hand, must imply a positive-unit flow deficit at its
node of origin ($C_{S_d}^d=1$), a negative-unit flow deficit at its destination
node ($C_{T_d}^d=-1$), and no flow deficit at any other node. Additionally,
Constraints~(\ref{C3}) and~(Cx) work, respectively, to ensure that no demand
uses more than $R_\mathrm{max}$ regenerators and that, taken together, the
demands using link $e$ ($d$ such that $x_p^dY_p^e=1$ for some $p\in P$) use no
more than the link's available capacity (LC).

Contemplating the CC requirements depends on one further set of variables, now
taking values from $\{0,1,\ldots,\text{LC}\}$. For $d\in D$ and $e\in E$, the
new variable is $z_d^e$ and serves to indicate, if greater than zero, the index
of the first FS used by demand $d$ on link $e$. To use these variables more
cleanly in the above formulation, it has once again proven convenient to adopt
the following additional shorthands:
\begin{align}
X_d^e &=
1 \text{ if }\sum_{p\in P}x_p^dY_p^e>0; \text{ }0 \text{ otherwise},
\nonumber\\
O_{d,d'}^e &=
1 \text{ if }z_d^e\le z_{d'}^e\le z_d^e+F_d^e-1; \text{ }0 \text{ otherwise}.
\nonumber
\end{align}
$X_d^e$ indicates whether demand $d$ uses link $e$ and $O_{d,d'}^e$ indicates
whether at least one of link $e$'s FSs is used by more than one demand. Clearly,
$X_d^e=0$ if and only if $F_d^e=0$. The shorthands $X_d^e$ and $O_{d,d'}^e$ can
be used to constrain the values of the $z_d^e$'s so that the CC requirement is
enforced.

Constraints~(\ref{C4})--(\ref{C7}) are now needed. Constraint~(\ref{C4}) is used
to ensure that $z_d^e$ is not a valid FS index ($z_d^e=0$) when demand $d$ does
not use link $e$ ($X_d^e=0$). If demand $d$ does use link $e$ ($X_d^e=1$), then
Constraint~(\ref{C5}) ensures that the value of $z_d^e$ is such that the indices
of the remaining $F_d^e-X_d^e=F_d^e-1$ FSs are no higher than LC. When demand
$d$ uses link $e\neq I_p$ on segment $p$ ($X_d^eY_p^e=X_d^{I_p}Y_p^{I_p}=1$),
Constraint~(\ref{C6}) ensures continuity by enforcing $z_d^e=z_d^{I_p}$.
Contiguity is ensured by Constraint~(\ref{C7}), according to which no two
distinct demands $d,d'$ can be such that
$z_{d'}^e\in\{z_d^e,\ldots,z_d^e+F_d^e-1\}$ for any link $e$.
Constraints~(\ref{C4})--(\ref{C7}) subsume Constraint~(Cx), which is therefore
not part of the formulation.

\section{RMSA-BP Formulation, Revisited}
\label{sec:form2}

The formulation given in Section~\ref{sec:form1} is correct in terms of
reflecting our understanding of the problem's three concomitant objectives, and
also in terms of laying down the constraints that guide the assignment of values
to its variables. However, in preliminary experiments it proved excessively
time-consuming even in relatively simple cases, owing mainly to the need to
comply with the CC requirement, that is, the need to satisfy
Constraints~(\ref{C4})--(\ref{C7}). In this section we describe two alterations
to the formulation that ended up making considerable difference in terms of
performance. Together with some fine-tuning of the solver employed, to be
described in Section~\ref{sec:results}, these alterations have made it possible
to enlarge the range of exactly solvable instances quite widely.

The first alteration is to incorporate a preprocessing step (i.e., a step prior
to calling the solver). This step creates, for each $d\in D$, a set $S_d$
comprising all solutions that admit demand $d$ while complying with
Constraints~(\ref{C2}), (\ref{C3}), and~(Cx). Each solution $s\in S_d$ is
therefore an assignment of values to the $x_p^d$'s that fully comply with the
flow-conservation requirement in Constraint~(\ref{C2}) for $a_d=1$, as well as
with the upper bounds $R_\mathrm{max}$ and LC imposed by Constraints~(\ref{C3})
and~(Cx), respectively. Such explicit enumeration is in general out of the
question, but in the case at hand it has proven feasible for many instances of
the problem. This is owed mainly to the pruning effect of $R_\mathrm{max}$,
which as mentioned in Section~\ref{sec:intro} is in practice already assigned a
small value.

For $s\in S_d$, we denote the value assigned to each $x_p^d$ during
preprocessing by $x_p^d(s)$. These values give rise to the following  useful
constants:
\begin{align}
R_d^s &= \sum_{p\in P}x_p^d(s)-1,
\nonumber\\
F_d^s &= \sum_{e\in E}\sum_{p\in P}F_p^dx_p^d(s)Y_p^e,
\nonumber\\
X_d^{e,s} &=
1 \text{ if }\sum_{p\in P}x_p^d(s)Y_p^e>0; \text{ }0 \text{ otherwise}.
\nonumber
\end{align}
$R_d^s$ is the number of regenerators used by demand $d$ in solution $s$,
$F_d^s$ is the total number of FSs used by demand $d$ in solution $s$, and
$X_d^{e,s}$ indicates whether demand $d$ uses link $e$ in solution $s$. The
revised RMSA-BP formulation we present uses three sets of variables: $x_d^s$, a
binary variable for each $d\in D$ and each $s\in S_d$, indicating whether demand
$d$ uses solution $s$; the $z_d^e$'s already used in Section~\ref{sec:form1};
and the binary variables $o_{(d,d')}^e$ and $o_{(d',d)}^e$, for each unordered
pair $(d,d')$ of distinct demands from $D$ and each link $e\in E$, used in
simplifying the enforcement of the contiguity part of the CC requirement. We
denote the set of such unordered pairs by $D_\mathrm{u}^2$. The formulation also
uses the shorthand
\begin{equation}
A_d=\sum_{s\in S_d}x_d^s.
\nonumber\\
\end{equation}
$A_d$ is the number of solutions in $S_d$ that demand $d$ uses and, provided
$x_d^s=1$ for at most one solution $s\in S_d$, indicates whether demand $d$ is
admitted. The objective function in the first formulation can thus be rewritten
as
\begin{equation}
\varphi=\sum_{d\in D}\Bigl(
w_1A_d-
w_2\sum_{s\in S_d}R_d^sx_d^s-
w_3\sum_{s\in S_d}F_d^sx_d^s
\Bigr).
\nonumber\\
\end{equation}
The revised RMSA-BP formulation is as follows.
\begin{align}
\text{max }
& \varphi
\nonumber\\
\text{s.t. }
\nonumber\\
& \sum_{s\in S_d}x_d^s\le 1
& \forall d\in D \tag{R1}\label{R1} \\
& (X_d^{e,s}Y_p^ez_d^e
& \forall d\in D, p\in P, \tag{R2}\label{R2} \\
& -X_d^{I_p,s}Y_p^{I_p}z_d^{I_p})x_d^s=0
& e\in p, e\neq I_p,
\nonumber\\
&& s\in S_d
\nonumber\\
& o_{(d,d')}^e+o_{(d',d)}^e\le 1
& \forall (d,d')\in D_\mathrm{u}^2, \tag{R3}\label{R3} \\
&& e\in E
\nonumber\\
& z_d^e+F_d^e\le z_{d'}^e+Mo_{(d,d')}^e
& \forall (d,d')\in D_\mathrm{u}^2, \tag{R4}\label{R4} \\
&& e\in E
\nonumber\\
& z_{d'}^e+F_{d'}^e\le z_d^e+Mo_{(d',d)}^e
& \forall (d,d')\in D_\mathrm{u}^2, \tag{R5}\label{R5} \\
&& e\in E
\nonumber
\end{align}

Constraint~(\ref{R1}) ensures that no demand $d$ uses more than one solution
from $S_d$. Constraints~(\ref{R2})--(\ref{R5}) target the CC requirement, with
Constraint~(\ref{R2}) ensuring that continuity holds and the remaining three
taking care of contiguity. Constraints~(\ref{R3})--(\ref{R5}) implement the
second performance-oriented alteration mentioned earlier in this section. The
reason why ensuring contiguity has such negative impact on the performance of
the first RMSA-BP formulation is the roundabout way Constraint~(\ref{C7})
approaches it. A much more direct way would be to do something in the style of
the $M$-independent part of Constraints~(\ref{R4}) and~(\ref{R5}), viz.,
$z_d^e+F_ e^d\le z_{d'}^e$ and $z_{d'}^e+F_{d'}^e\le z_d^e$ for each unordered
pair $(d,d')$ of demands. These, however, can never be concomitantly satisfied,
essentially forcing an a priori choice between them. This impossibility is what
the $M$-dependent part of the two constraints helps circumvent. For a
sufficiently large value of $M$, and given that Constraint~(\ref{R3}) disallows
the occurrence of $o_{(d,d')}^e=o_{(d',d)}^e=1$, it is possible, e.g., that
Constraint~(\ref{R4}) ends up enforcing $z_d^e+F_ e^d\le z_{d'}^e$
(with $o_{(d,d')}^e=0$) while Constraint~(\ref{R5}) enforces
$z_{d'}^e+F_{d'}^e\le z_d^e+M$ ($o_{(d',d)}^e=1$), avoiding the superposition of
the two demands' FSs while opening up the possibility of $z_d^e<z_{d'}^e$.
Clearly, to ensure that the two constraints function in this way it suffices
that we have $M>\text{LC}$, so we use $M=\text{LC}+1$ throughout. The overall
strategy is known as the bigM approach to handle indicator-dependent constraints
efficiently \cite{bigM}.

\section{Computational Results and Conclusion}
\label{sec:results}

Our computational experiments involved solving RMSA-BP, as formulated in
Section~\ref{sec:form2}, on a computer with two AMD EPYC 7763 64-core processors
and 512 GB RAM. We used the Gurobi 9.5.2 solver along with the NetworkX 3.1
Python package on the Debian 11 operating system. All the RMSA-BP instances we
considered are relative to a network with the NSFNET topology and modulation,
per-slot efficiency, and maximum reach as in \cite{turco}. The experiment for
each instance consisted in preprocessing for solution enumeration followed by
optimization, and was allowed to run for no more than 30 hours before timeout.
Gurobi was allowed no more than 32 parallel threads per instance.

Our use of the Gurobi solver was predicated on the adoption of three of the
features it offers that turned out to be of crucial importance. First, to
address the prioritization symbolized by weights $w_1,w_2,w_3$ in our
formulation's objective function, we used the solver's multi-objective
hierarchical mode, which allows a priority to be set for each individual
objective and ensures that they are optimized from highest to lowest priority.
Solving for an objective does not affect any solution found when solving for
higher-priority objectives. Second, to ensure that the problem's formulation is
indeed of the ILP type, we used the solver's indicator-type constraints. These
are instrumental in view of the otherwise quadratic nature of Constraint~(R2).
Third, we used the solver's parameter tuning tool (grbtune) prior to any actual
experiment. This tool is given one or more instances to solve and analyzes them
automatically to fine-tune the solver's internal parameters. Relying on the
parameter values it outputs is based on the expectation that they will likewise
lead to good performance on the instances it has not analyzed. We found this to
be generally true.

\begin{table}[t]
\setlength{\tabcolsep}{5pt}
\centering
\caption{Average Blocking and Usage Results, 30 Instances}
\label{tab:tabresult}
\begin{tabular}{c|cccc|cccc}
\hline
$\vert D\vert$ & $R_\mathrm{max}$ & BD & TR & TFS & $R_\mathrm{max}$ & BD & TR & TFS \\
\hline
100 &  1  & 1.0 & 43.4 & 1\,265 &  2  & 0.6 & 42.4 & 1\,263  \\
110 &  1  & 4.4 & 55.1 & 1\,292 &  2  & 2.9 & 59.0 & 1\,283  \\
120 &  1  & 7.5 & 55.7 & 1\,329 &  2  & 5.1 & 64.8 & 1\,319  \\
\hline
\end{tabular}

\end{table}

\begin{table}[t]
\setlength{\tabcolsep}{5pt}
\centering
\caption{Timeouts and Execution Times (h:m:s), 30 Instances}
\label{tab:ttimes}
\begin{tabular}{ccccccc}
\hline
$\vert D\vert$ & $R_\mathrm{max}$ &  TO & Minimum & Maximum & 80th Percentile \\
\hline
100 & 1 & 2 & 00:03:06 & 15:48:10 & 00:50:44 \\
110 & 1 & 3 & 00:04:15 & 14:03:06 & 01:28:03 \\
120 & 1 & 8 & 00:14:24 & 14:24:14 & 01:57:48 \\
\hline
100 & 2 & 4 & 00:07:52 & 29:39:28 & 04:42:32 \\
110 & 2 & 1 & 00:03:39 & 12:27:29 & 03:51:46 \\
120 & 2 & 4 & 00:08:26 & 14:03:24 & 08:34:09 \\
\hline
\end{tabular}

\end{table}

\begin{figure}[t]
\centering
\includegraphics[width=1\linewidth]{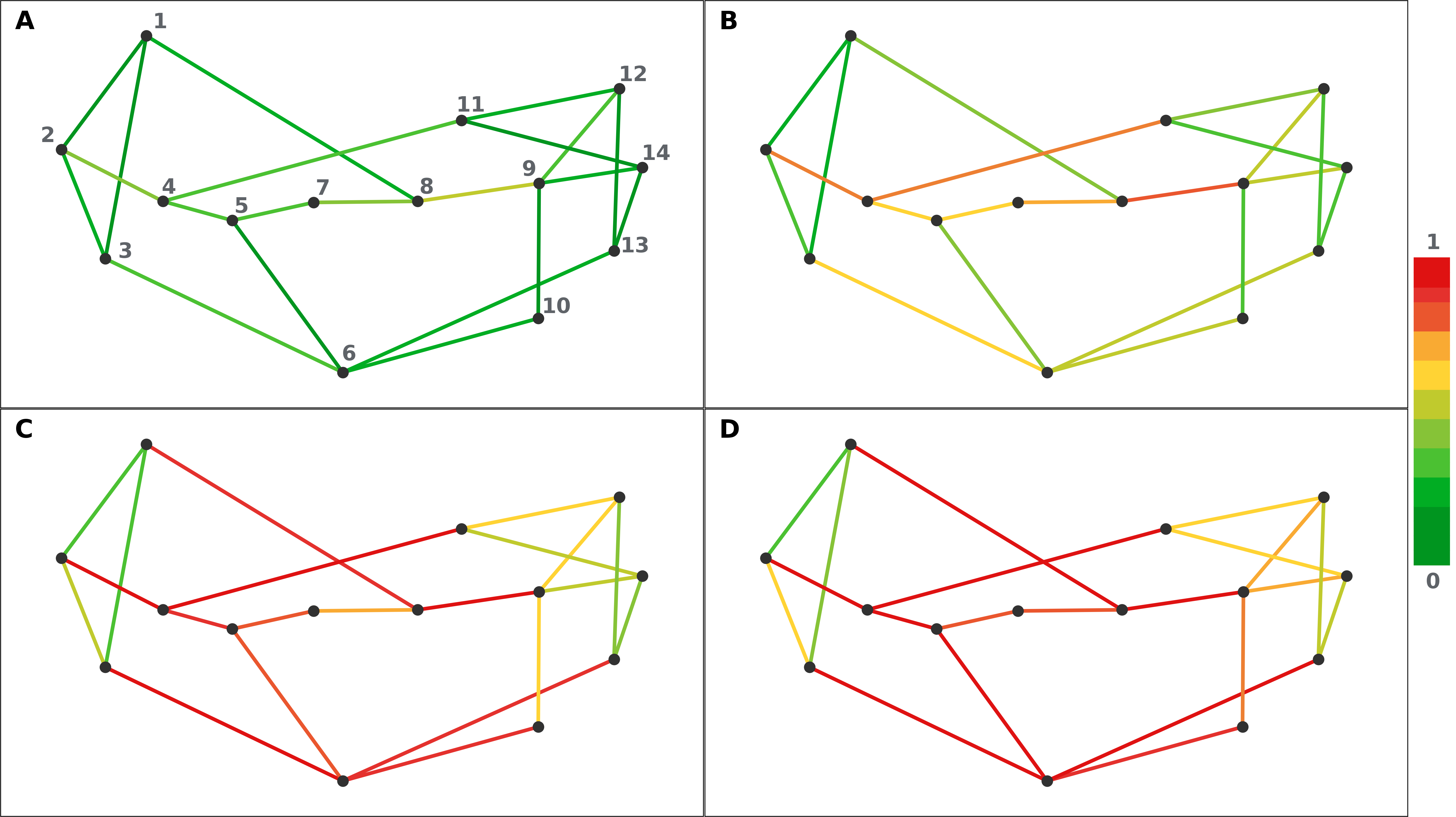}
\caption{Link-usage heat map for $\vert D\vert=30$ (A), 60 (B), 90 (C), 120
(D).}
\label{fig:heat-map}
\end{figure}

We used 90 instances for each value of $R_\mathrm{max}$, 30 of them with
$\vert D\vert =100$, 30 with $\vert D\vert=110$, and 30 with $\vert D\vert=120$.
Each demand $d$ had bandwidth $B_d=100$~Gbps, and nodes of origin $S_d$ and
destination $T_d$ chosen uniformly at random. We initially used $\text{LC}=160$
but this never resulted in any demand being blocked. We then turned to
$\text{LC}=80$, which made it possible for links to saturate and more
interesting results to be observed. We report on the $\text{LC}=80$ cases
exclusively.

For $R_\mathrm{max}=1,2$, in Table~\ref{tab:tabresult} we show the average
number of blocked demands ($\text{BD}\le\vert D\vert$), the average total
number of regenerators used ($\text{TR}\le\vert D\vert\,R_\mathrm{max}$), and
the average total number of FSs used per demand
($\text{TFS}\le 21\,\text{LC}=1\,680$, since the NSFNET has 21 links). In
Table~\ref{tab:ttimes}, we see that the number of timeouts ($\text{TO}\le 30$)
was consistently low, however with considerable variation in execution times for
successful instances. This notwithstanding, for 24 (80\%) of the instances they
fell below a small number of hours, though substantially more for
$R_\mathrm{max}=2$ than for $R_\mathrm{max}=1$. Supplementing the information in
the tables, we note that the average numbers of CPU cores used were between 20
and 22.

We also conducted one further set of experiments to help in understanding how
network topology influences link usage as the number of demands increases. The
setup is still mostly the same as in the previous experiments, except that now
four demand sets are used, with $\vert D\vert=30,60,90,120$. A link's usage is
the ratio of the number of FSs used on it to LC. A heat map with averages over
each demand set's 30 instances is shown in Figure~\ref{fig:heat-map}. Even
though the increase in link usage does, as expected, grow with $\vert D\vert$,
somewhat unexpectedly we also see that link $(8,9)$ is always one of the most
used in all cases, while $(1,2)$ is one of the least used. These seem like
inherent structural properties of the NSFNET topology, which in principle might
not have come up if some heuristic had been used instead. To conclude, we then
note that therein lies the importance of exact approaches like RMSA-BP, since
they can provide crucial aid in the analysis of network topology and usage, and
through such analysis can influence design and deployment policies. Further
research should concentrate on generalizing the objective function we have used
to meet other goals, as well as on seeking additional improvement opportunities
to both problem formulation and solver-feature exploitation, so that larger
networks can be handled as well.

\bibliographystyle{IEEEtran}
\bibliography{rmsabp-abbrev}

\end{document}